\documentclass[pra,superscriptaddress, showpacs,  twocolumn]{revtex4}
\usepackage{graphicx}       
\usepackage{bm}         
\usepackage{amsmath}        
\usepackage{latexsym}
\newcommand{\beq}{\begin{equation}}

\newcommand{\bb}{\langle\hat B\rangle}
\newcommand{\cc}{\langle\hat A\rangle}

\begin{document}

\title{Estimating the expectation values of spin 1/2 observables with finite resources}
\author{Thomas Brougham and Erika Andersson}
\affiliation{SUPA, Department of Physics, University of Strathclyde, Glasgow G4
ONG, UK}

\begin{abstract}
We examine the problem of estimating the expectation values of two observables when we have a finite number of copies of an unknown qubit state.  Specifically we examine whether it is better to
measure each of the observables separately on different copies or to perform a joint measurement of the observables on each copy.  We find that joint measurements can sometimes provide an advantage over
separate measurements, but only if we make estimates of an observable based solely on the results of measurements of that observable.  If we instead use both sets of results to estimate each observable then we find that individual 
measurements will be better.  Finally we consider estimating the expectation values of three complementary observables
for an unknown qubit.
\end{abstract}
\maketitle

\section{Introduction}
One of the most fundamental questions in physics is how to best determine unknown properties of a physical system.  In quantum mechanics this problem is complicated by the
probabilistic nature of the theory.  The expectation value of an observable is, however, well defined in quantum mechanics.  The conceptual problems are intensified when we seek to obtain information about two observable quantities.  Classically 
the only limit to how well we can measure two quantities is the skill of the experimenter and the available technology.  In quantum mechanics, however, there exist fundamental limits on our ability
to measure two non-commuting observables.  It is sometimes thought that it is impossible to simultaneously measure two non-commuting observables upon the same system. 
This, however, is
not the case; such joint measurements are possible at the expense of increased uncertainty in the measurement results \cite{AK, AG, opQM}.

One of the earliest investigations into estimating parameters of a quantum system was by Helstrom \cite{helstrom}.  Parameter estimation is related to state estimation, for if we
determine enough parameters then it is possible to completely specify the state of the system.  There exists a large literature on quantum state estimation and
quantum state tomography, see for example \cite{tomorev, tomoexper, gentheory, biasedtomo, stateestimate}.  One important consideration for state tomography is the available number of copies of the 
undetermined state.  When we have a very large number of copies then the error in determining the state will be small for all sensible protocols.  If, however, there is a limited
number of copies available then it becomes a priority to find efficient means of estimating the state \cite{thesis}.  Similar considerations are important when
trying to estimate the expectation value of an observable.  Methods that prove reliable when we have a very large number of copies of a state may not be the most efficient
approach when the number of copies is smaller.

The problem we shall address in this paper is the following.  Suppose that we have a finite number of copies of some unknown state of a two-level system.  Our task is to 
estimate the expectation values of two different observables, while minimizing the total error in the estimates.  For the case of estimating a single observable, it was found
that performing a collective measurement on all the copies yields no advantage over separate measurements upon the individual copies of the system \cite{DAriano,
Hayashi}. When we consider two observables, the
added complexity may mean that collective measurement could provide an advantage.  In this paper, however, we
shall not consider collective measurements but will instead focus on simpler measurement schemes that are also more amenable to experimental realisation.  Our aim is to determine whether it is better to measure a
single observable upon each system or to perform a joint measurement of two or three observables upon each system.  We shall find that if each expectation value is estimated using only
the results for that
particular observable, then sometimes a joint measurement does provide an advantage.  This improvement, however, is due to the simple way the expectation values have been
estimated.  By using all the measurement results it is possible to improve the performance for the case when we measure each observable on a separate system. 

The paper will be organized in the following way.  A brief review of joint measurements will be given in section \ref{joint}.  In section \ref{separate} we will examine
making separate measurements of the observables, before considering joint measurements.  We will then investigate whether biasing the results of the measurements can improve the
performance, both for separate measurements and for joint measurements.  In section \ref{bayes} we ask whether Bayes' rule can be used to make better use of the measurement results when we are
performing individual measurements of each observable.  In section \ref{3obs} we
shall briefly investigate estimating the $x$, $y$ and $z$ components of spin for a spin-1/2 particle.  Finally we discuss our results in section \ref{conc}.

\section{Joint measurements of spin 1/2 observables}
\label{joint}
A joint measurement of two observables is a simultaneous measurement of both observables upon the same quantum system.  When the observables of interest commute then a joint measurement can be accomplished
with standard von Neumann or projective quantum measurements.  If the observables do not commute, then we must describe our measurements in terms of the probability
operator measure (POM) formalism.  A detailed description of this generalized description of measurements can be found in \cite{nch, peres}.  

A condition that is often used for joint measurements is the joint unbiasedness condition, which requires that the expectation values for the jointly measured
observables are proportional to the expectation values of the observables measured separately \cite{AG, opQM}.  We thus require that $\langle\hat A_{J}\rangle=\alpha\langle\hat A\rangle$ and $\langle\hat B_{J}\rangle=\beta\langle\hat
B\rangle$, where $\langle\hat A_{J}\rangle$ and $\langle\hat B_{J}\rangle$ are the jointly measured expectation values and $\alpha$ and $\beta$ are constants of
proportionality.  For examples of the joint unbiasedness condition applied to spin-1/2 systems see \cite{busch, busch2, demuynck, steves, erikas}.  It is possible to relax the
condition of joint unbiasedness.  This leads to a more general description of joint measurements \cite{hall}.    

For spin-1/2 observables, $\hat A={\bf a\cdot}\boldsymbol{\hat\sigma}$ and $\hat B={\bf
b\cdot}\boldsymbol{\hat\sigma}$, the eigenvalues are $\pm 1$.  We will choose to assign the numerical values of $\pm 1$ to the results spin up and down also for a joint measurement of $\hat A$
and $\hat B$.  This means that $|\alpha|$ and $|\beta|$ will vary between one and zero.  We wish the joint measurements to represent, as accurately as
possible, a measurement of both $\hat A$ and $\hat B$.  We thus want the marginal probability distributions for the joint and the separate measurements to be as
similar as possible.  For
spin $1/2$ observables the probabilities are exactly specified by the expectation values, hence we wish to make the expectation values of the jointly measured observables as
close as possible to the expectation values of the separately measured observables.  This is achieved by making $\alpha$ and $\beta$ as close to one as is
possible.  The values of $\alpha$ and $\beta$ will be restricted by the inequality \cite{busch}
\begin{equation}
\label{ineq}
 |\alpha{\bf a}+\beta{\bf b}|+|\alpha{\bf a}-\beta{\bf b}|\le 2.
\end{equation}
If a joint measurement scheme allows us to saturate inequality (\ref{ineq}), then, for given directions {\bf a} and {\bf b}, this measurement gives 
the largest possible value of  $|\alpha|$ for a given $\beta$ and vice versa.  Any joint measurement of two components of spin  for which inequality  
(\ref{ineq}) is saturated is in this sense an optimal joint measurement.  For a description of how optimal joint measurements of spin can be implemented see \cite{busch2, demuynck, erikas, cloning}.

\section{Estimation of two observables}
\label{separate}
Suppose we have $2N$ copies of some unknown pure qubit state and we wish to learn the expectation values of the spin observables $\hat A={\bf a}\cdot\boldsymbol{ \hat \sigma}$ and $\hat
B={\bf b}\cdot\boldsymbol{ \hat \sigma}$, where ${\bf a}$ and ${\bf b}$ are unit vectors.  The most general approach would be to perform a collective measurement on all the $2N$ copies.  When measuring a single
observable it has been shown, however, that separate measurements on each quantum system are optimal \cite{DAriano, Hayashi}.  For the case of two observables, the added
complexity may mean that performing a collective measurement could provide a benefit.  It may, however, be difficult to implement collective measurements.  Instead we shall
investigate theoretically simpler methods, which will also be more experimentally amenable.  

The first strategy we shall investigate is to perform measurements of $\hat A$ on $N$ of the copies and of $\hat B$ on
the remaining $N$ copies.  The second strategy is to perform a joint measurement of both $\hat A$ and $\hat B$ on all $2N$ copies of the unknown state.  As a figure of merit we shall use the averaged
square error.  The results of each measurement will be recorded.  After all the measurements have been performed, we will be left with a set of either $2N$ or $4N$ outcomes, depending on which of the two
measurement strategies we were using.  The estimates of $\langle\hat A\rangle$ and $\langle\hat
B\rangle$ will depend upon which of the $2^{2N}$ or $2^{4N}$ possible sets of measurement results was obtained.  We shall denote our estimates of $\langle\hat A\rangle$ and $\langle\hat B\rangle$ as $\tilde a_j$ and
$\tilde b_k$ respectively.  The subscripts $j$ and $k$ are used to indicate which of the possible sets of outcomes has been used to calculate the estimates.  The square of the
error in the estimates of $\langle\hat A\rangle$ and $\langle\hat B\rangle$ can now be defined as 
\begin{equation}
\label{errorsAB}
\epsilon^a_j=(\langle\hat A\rangle -{\tilde a_j})^2\text{  and  }\epsilon^b_k=(\langle\hat B\rangle -{\tilde b_k})^2,
\end{equation}
respectively. For different sets of results we will obtain different estimates and
thus we will obtain different errors.  Hence we shall average the error over all possible sets of outcomes.  Let $P_j$ denote the probability of obtaining the $j^{th}$ set of outcomes for our measurements of the observables upon the $2N$
copies, given that we have a particular state $\hat\rho$.  Then the total averaged square error is given by
\begin{eqnarray}
\label{totalerror}
\epsilon_T&=&\int\sum_{j}{P_{j}(\langle\hat A\rangle-{\tilde a}_j)^2}d\hat\rho\nonumber\\
&+&\int\sum_{j}{P_{j}(\langle\hat B\rangle-{\tilde b}_j)^2}d\hat\rho,
\end{eqnarray}
where the integral $\int ...d\hat\rho$ indicates an average over all pure states.  To perform this average we will assume that all the pure states are evenly distributed about
the surface of the Bloch sphere.  We can introduce $\theta$ as the angle between the $z$ axis of the Bloch sphere and the Bloch vector of the pure state.  We also introduce
$\phi$ as the angle between the $x$ axis and the projection of the Bloch vector of the state in the $xy$ plane of the Bloch sphere.  The average $\int{...d\hat\rho}$ can thus be
re-expressed as $1/4\pi\int^{\pi}_0{\int^{2\pi}_0{...\sin(\theta)d\theta}d\phi}$.   

Consider now the first approach where we measure only one observable on each copy of the state.  The choice of estimate is important.  For
now we shall simply take the mean of the measurement results as our estimates, i.e. $\tilde{a}_j=\sum_i{a_j^i}/N$ and $\tilde{b}_j=\sum_i{b_j^i}/N$, where $a_j^i$ and $b_j^i$ are the $i^{th}$
measurement results from the $j^{th}$ set of measurement outcomes of $\hat A$ and $\hat B$ respectively.  The outcomes of a measurement will be either $+1$ or $-1$.  The error in $\langle\hat
A\rangle$ will be given by 
\begin{eqnarray}
\label{errorA}
\epsilon_a&=&\int\sum_{j}{P_{j}(\langle\hat A\rangle-{\tilde a}_j)^2}d\hat\rho\nonumber\\
&=& \int \text{var}({\tilde a})+(\langle\hat A\rangle-\langle {\tilde a}\rangle)^2d\hat\rho,
\end{eqnarray}
where var$({\tilde a})=\sum_j{P_j({\tilde a_j}-\langle{\tilde a}\rangle)^2}$ is the variance in $\tilde{a}_j$ and $\langle \tilde{a}\rangle=\sum_{j}{P_{j}\tilde{a}_j}$.  It can easily be seen that
$\langle\tilde{a}\rangle=1/N\sum_i{\langle
a^i\rangle}=\langle\hat A\rangle$ and thus this choice of estimate is unbiased.  Hence the only term contributing to the error in equation (\ref{errorA}) is the variance of ${\tilde a}_j$.  Using the definition of
variance it is straightforward to show that 
\begin{eqnarray}
\text{var}({\tilde a})=\frac{1}{N^2}\sum_j{P_j \left[\sum_m{(a_j^m)^2}+\sum_{m, n(m\ne n)}{a_j^ma^n_j}\right]}\nonumber\\
-\frac{1}{N^2}\left[\sum_m{\langle a^m\rangle^2}+\sum_{m, n (m\ne n)}{\langle a^m\rangle\langle
a^n\rangle}\right].
\end{eqnarray}
The results of each measurement are independent, which means that $\sum_j{P_ja^m_ja^n_j}=\langle a^ma^n\rangle=\langle a^m\rangle\langle
a^n\rangle$.  It is also useful to note that $(a_j^m)^2=1$.  From these observations it follows that var$({\tilde a})=(1-\langle\hat A\rangle^2)/N$.  The error is obtained by averaging over all pure
states.  If we assume that all the pure states are equally likely then we find that the average of $\langle\hat A\rangle^2=\langle {\bf
a\cdot}\boldsymbol{\hat\sigma}\rangle^2=1/3$, and thus $\epsilon_a=2/(3N)$.  The calculation of the error in the estimate of $\langle\hat B\rangle$ proceeds in a similar way and leads to the total
error being
\begin{equation}
\label{error1}
\epsilon_T=\frac{4}{3N}.
\end{equation}  

Equation (\ref{error1}) was derived for the situation where each observable is measured on $N$ copies of the unknown state.  Instead we could measure $\hat A$ on $N_1$ copies of the
state and measure $\hat B$ on the remaining $2N-N_1$ copies.  It can be shown that the total error for this case would be $\epsilon_T=2/(3N_1)+2/(3[2N-N_1])$.  A simple calculation shows that
$\epsilon_T$ has a minimum when $N_1=N$, which leads to the error being given by equation (\ref{error1}). 

\begin{figure}
\center{\includegraphics[width=8cm,height=!]
{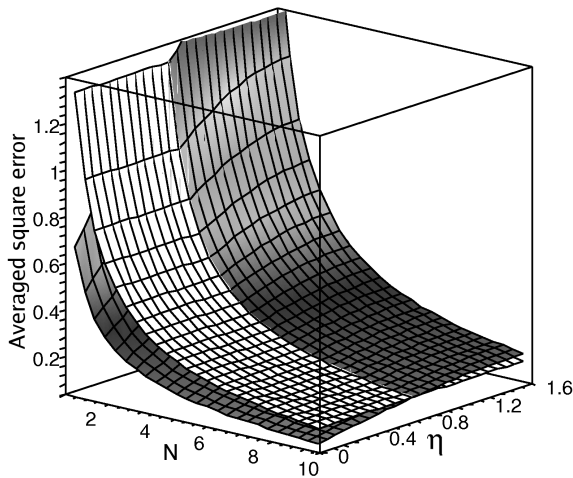}}
\caption{A plot of the error as a function of $\eta$, the angle between ${\bf a}$ and ${\bf b}$, and $N$, which is half the total number of copies of the unknown state.  The
white surface represents the error for separate measurements of $\hat A$ and $\hat B$, given by equation (\ref{error1}).  The gray
surface represents the error for joint measurements, given by equation (\ref{JM}).}
\label{fig1}
\end{figure}

We will now look at the second strategy, where we perform joint measurements of both $\hat A$ and $\hat B$ on each of the $2N$ copies of the unknown state.  An obvious choice of
estimates would be $\tilde{a}_j=\sum_i{{a_j^i}}/(2N)$ and $\tilde{b}_j=\sum_i{{b_j^i}}/(2N)$.  From the joint unbiasedness condition, however, we know that the expectation
values of the jointly measured observables would not be equal 
to $\langle\hat A\rangle$ and $\langle\hat B\rangle$.  Using these estimates would therefore lead to an error that would never tend to zero.  Instead we will use the estimates
$\tilde{a}_j=\sum{a_j^i/(\alpha2N)}$ and  $\tilde{b}_j=\sum{b_j^i/(\beta2N)}$.  This is equivalent to relabeling our measurement outcomes $\pm 1/\alpha$ and $\pm 1/\beta$ for a measurement of $\hat A$
and $\hat B$ respectively.  By following an argument similar to that used to derive equation (\ref{error1}) it is possible to show that the total error takes the form
\begin{equation}
\label{errorJM}
\epsilon_T=\frac{1}{2N}\left(\frac{1}{\alpha^2}+\frac{1}{\beta^2}-\langle\hat A\rangle^2-\langle\hat B\rangle^2\right).
\end{equation} 
The symmetry in the problem suggests that we will obtain a minimum for (\ref{errorJM}) when $\alpha=\beta$, that is, when we measure both observables equally well.  This can be verified by minimizing
(\ref{errorJM}) subject to the constraint imposed by (\ref{ineq}).  By averaging over all pure state we find that the total error is
\begin{equation}
\label{JM}
\epsilon_T=\frac{1}{N}\left(\frac{1}{\alpha^2}-\frac{1}{3}\right).
\end{equation}
If we solve (\ref{ineq}) for $\alpha=\beta$, we find that $\alpha=\sqrt{1/(1+|\sin(\eta)|)}$, where $\eta$ is the angle between ${\bf a}$ and ${\bf b}$.  It
can be seen that equation (\ref{JM}) has the correct asymptotic behavior as for $N\rightarrow\infty$ we find that $\epsilon_T\rightarrow 0$.  A plot of the error functions, equation
(\ref{error1}) and equation (\ref{JM}), is
shown in figure \ref{fig1}.  It can be seen that when the angle between ${\bf a}$ and ${\bf b}$ is small, joint measurements can provide a total error that is lower than in
equation (\ref{error1}).  As the angle
increases it becomes more advantageous to perform separate measurements of the observables.  It can be shown that joint measurements are better than separate measurements when
$|\sin(\eta)|\le 2/3$.    

When the angle between ${\bf a}$ and ${\bf b}$ is small then in a sense the observables $\hat A$ and $\hat B$ can be thought to share information.  A joint measurement of $\hat A$ and $\hat B$ is
naturally able to exploit the shared information.  This is because according to equation (\ref{ineq}), when the angle between ${\bf a}$ and ${\bf b}$ is small, $\alpha$ and $\beta$ can be close to 1,
corresponding to sharper measurements of $\hat A$ and $\hat B$. This also helps to explain why separate
measurements are more effective when the angle between ${\bf a}$ and ${\bf b}$ gets closer to $\pi/2$, since at this angle the observables are complementary.  As $\eta$ tends to
$\pi/2$ the two observables share less information meaning that joint measurements lose their advantage.  This suggests that we could try to
improve the performance of separate measurements by making our estimates using the measurement results of both $\hat A$ and $\hat B$.  This idea will be discussed in sections
\ref{biasing} and \ref{bayes}.  Before this we will discuss a simpler approach to decreasing the error in our estimates.

\section{Biased estimates}   
\label{biasing}
The estimates that were looked at in the previous section were unbiased.  This means that $\langle\tilde a_j\rangle=\sum_j{P_j\tilde a_j}=\langle\hat A\rangle$ and $\langle\tilde
b_j\rangle=\sum_j{P_j\tilde b_j}=\langle\hat B\rangle$.  In \cite{Hayashi}, it was found that adding bias could yield a lower error when estimating the
expectation value of a single observable.  This raises the possibility that biased estimates could lower the error in our estimates of the two expectation values.  When adding
bias, care must be taken in order to obtain the correct asymptotic behavior.  As $N$ tends to infinity, the error should tend to zero.  In order for this to
occur the bias must tend to zero as $N$ tends to infinity.  We will now look at adding biasing to both of the measurement strategies that we discussed in section \ref{separate}.

Consider now the first measurement strategy, where we measured $\hat A$ and $\hat B$ on separate copies of the unknown system.  We shall bias the results of the measurement of
$\hat A$ by
rescaling the measurement outcomes from $\pm 1$ to $\pm K$.  This is equivalent to take our estimate  of $\langle\hat A\rangle$ to be
$\tilde{a}=\sum_i{Ka^i}/N$.  By following a modified form of the argument that led to equation (\ref{error1}) it can be found that the error in $\hat A$ is now 
\begin{equation}
\epsilon_a=\frac{(K-1)^2}{3}+\frac{2K^2}{3N}.
\end{equation} 
It can easily be seen that the value of $K$ that minimizes the error is $K=N/(N+2)$, which leads to $\epsilon_a=2/(3(N+2))$ and to $\tilde{a}=\sum_i{a^i/(N+2)}$.  A similar
argument can be followed to show that the minimum error in $\langle\hat B\rangle$ is equal to $\epsilon_a$, which leads to 
\begin{equation}
\label{biased}
\epsilon_T=\frac{4}{3(N+2)}.
\end{equation}
It can be seen that equation (\ref{biased}) is always less than equation (\ref{error1}) for $N\ge 1$.  Adding bias has thus decreased the error in our estimate.  

To try to understand how
adding bias helps we shall look at the case when $N=1$.  In this situation our estimate will just be the value that we assign for the outcome of a single measurement.  When we
are not biasing the outcomes, then if we get the outcome spin up, our estimate would be $\tilde a=1$.  This corresponds to guessing that the initial state was the eigenstate of ${\bf
a\cdot\hat\sigma}$, with eigenvalue 1.  It is possible that our initial state was prepared as this state, however, it is more likely that this was not the case.  It is more
probable that $\langle\hat A\rangle$ is less than one, and thus it is preferable to take our estimate $\tilde a$ to be less than one.  If bias is added, then our
estimate, based on a single spin up result, will be 1/2.  Biasing thus stops us from overestimating the magnitude of the expectation value when $N$ is small.

Equation (\ref{biased}) was derived for the situation where $\hat A$ and $\hat B$ are each measured on $N$ copies of the unknown state.  Alternatively we could have measured $\hat A$ on
$N_1$ copies of the state and measured $\hat B$ on the remaining $2N-N_1$ copies.  It can be shown that the total error in this case is $\epsilon_T=2/[3(N_1+2)]+2/(3[2N-N_1+2)]$.  A
straightforward calculation shows that the value of $N_1$ that minimises the total error is $N_1=N$, and thus the minimum error will again be given by equation (\ref{biased}).

We can consider a more general situation where we assign $K_1$ and $K_2$ to the measurement results instead of $\pm K$, where $|K_1|$ need not equal $|K_2|$.  It can then be
shown that a necessary condition to obtain the minimum error is that $K_1+K_2=0$, and hence we would again obtain equation (\ref{biased}).  It is worth comparing this with the
results of \cite{Hayashi}, where it was found that the form of the estimates that minimized the error was $[\sum_i{a^i+\text{Tr}(\hat A)}]/(N+d)$, where $d$ is the dimension of
the Hilbert space.  In our qubit situation this would lead us to choose $\tilde{a}=\sum_i{a^i}/(N+2)$ as our estimates for $\hat A$ and $\tilde{b}=\sum_i{b^i}/(N+2)$ as our
estimate for $\hat B$, which is what we have found by allowing bias in our results.

We shall now investigate whether adding bias to a joint measurement  will lead to
better estimates.  It should be noted that biased in this sense is not the same as when we discussed the joint unbiasedness condition in
section \ref{joint}.  Instead we mean that the joint measurements will be such that $\sum_j{P_j\tilde a_j}\ne\cc$ and $\sum_j{P_j\tilde b_j}\ne\bb$.  Consider first the results
for the observable $\hat A$.  Previously we assigned the numerical values $\pm 1/\alpha$; one way of biasing the results would be 
to instead assign $\pm K/\alpha$ to the results.  The estimate of $\langle\hat A\rangle$ now becomes $\tilde{a}_j=\sum_i{Ka_j^i}/(\alpha 2N)$.  It can be
shown that the error for this estimate is
\begin{equation}
\label{biasA}
\epsilon_a=\frac{K^2}{2N}\left(\frac{1}{\alpha^2}-\frac{1}{3}\right)+\frac{(1-K)^2}{3}.
\end{equation}
The value of $K$ that minimizes the error, $\epsilon_a$, is $K=(2N\alpha^2)/(3-\alpha^2+2N\alpha^2)$.  A similar argument can be followed to find the minimum error in
$\langle\hat B\rangle$.  The symmetry of the problem suggests that we should measure $\hat A$ and $\hat B$ equally well, and thus $\alpha=\beta$.  This can be verified by
minimising the total error subject to the inequality (\ref{ineq}).  The error in
$\langle\hat B\rangle$ will thus equal $\epsilon_a$, which leads to the following expression for the total error   
\begin{equation}
\label{biasJM}
\epsilon_T=\frac{2(3-\alpha^2)}{3(3-\alpha^2+2N\alpha^2)}.
\end{equation}
We should note that although we are using biased estimates, the expectation value, $\langle\hat A_J\rangle$, is still proportional to $\cc$.  The joint measurement will thus
still satisfy the joint unbiasedness condition.

\begin{figure}
\center{\includegraphics[width=8cm,height=!]
{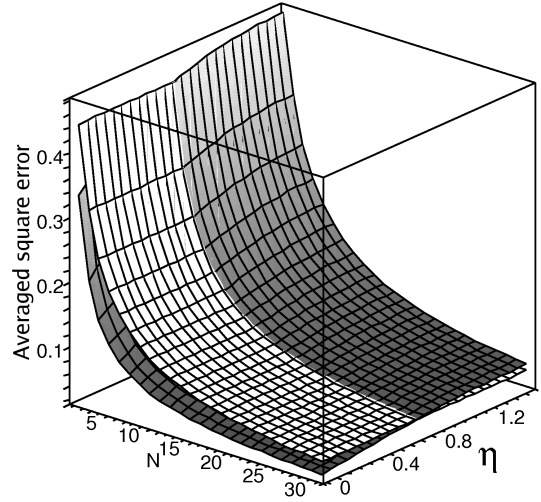}}
\caption{A plot of the error as a function of $\eta$, the angle between ${\bf a}$ and ${\bf b}$, and $N$ which is half the total number of copies of the unknown state.  The
white surface represents the error for biased separate measurements, given by equation (\ref{biased}).  The gray surface represents the error for biased joint
measurements, given by equation (\ref{biasJM}).}
\label{fig3}
\end{figure}

A more general way of biasing the results is to assign $K_1/\alpha$ and $K_2/\alpha$ to the results of the measurement of $\hat A$.  It can then be shown that a necessary
condition to minimize the error is that $K_1=-K_2$. This leads to the error in $\langle\hat A\rangle$ being equal to equation (\ref{biasA}) and thus the total error will again
be given by equation (\ref{biasJM}). 
By comparing equation (\ref{biasJM}) to equation (\ref{biased}), we find that biased joint measurements provide a smaller error than biased separate measurements for values of $\eta$ up to
$\sin^{-1}(2/3)\approx 0.73$ radians.  This is shown in figure \ref{fig3}.

When $\eta$ is small then $\hat A$ and $\hat B$ share information, which is exploited by joint measurements.  This is the reason why the error for joint measurements, equation
(\ref{biasJM}), is less than the error for separate measurement, equation (\ref{biased}), when $\eta$ is small.  If we wished to improve the performance of separate measurements
then we will need to make the estimates of $\cc$ and $\bb$ based on all the measurement results for both observables.  We will now investigate a simple way of achieving this. 

Suppose we perform separate measurements of the two observables and obtain the set of measurement outcomes $\{a^m_j, b^n_j\}$.  We shall take our estimate of $\cc$ to be
\begin{equation}
\label{Estsimple}
\tilde{a}_j=\frac{1}{N}\sum_m{Ka^m_j}+\frac{\lambda}{N}\sum_n{b^n_j},
\end{equation}
where $\lambda$ is a constant that weights the results for $\hat B$.  The constant $K$ is included to bias the results of $\hat A$.  This additional biasing may be needed as we
should expect that there will be situations where $\lambda=0$, for instance when ${\bf a}\cdot{\bf b}=0$.  It can be shown that 
\begin{eqnarray}
\epsilon_a&=&\int\text{var}(\tilde{a})+(\langle\hat
A\rangle-\langle\tilde{a}\rangle)^2d\hat\rho\nonumber\\
&=&\frac{2(K^2+\lambda^2)}{3N}+\frac{(1-K)^2+\lambda^2}{3}\nonumber\\
&-&\frac{2\lambda(1-K){\bf a}\cdot{\bf b}}{3}.
\end{eqnarray} 
The values of $K$ and $\lambda$ that minimize the error are given by $K=N[2+N-N({\bf a}\cdot{\bf b})^2]/[(N+2)^2-N^2({\bf a}\cdot{\bf b})^2]$ and
$\lambda=2N{\bf a}\cdot{\bf b}/[(2+N)^2-N^2({\bf a}\cdot{\bf b})^2]$.  The estimate of $\langle\hat B\rangle$ will take a form analogous to equation (\ref{Estsimple}).  This
means that the minimum error in the estimate of $\langle\hat B\rangle$ will equal the minimum of $\epsilon_a$.  The total error will thus be
\begin{equation}
\label{simple}
\epsilon_T=\frac{4[N({\bf a}\cdot{\bf b})^2-N-2]}{3[N^2({\bf a}\cdot{\bf b})^2-(N+2)^2]}.
\end{equation}
This new value of error is always less than the error found using biased joint measurements in equation (\ref{biasJM}) and is also less than equation (\ref{biased}).  A more
sophisticated means of estimating the expectation values using the results of all the measurements, will be discussed in the next section.

\section{Bayesian inference for two observables}
\label{bayes}
The Bayesian view of probability is that it encodes what we know about a system.  For example if someone flips a coin, we might say that there is a probability of $1/2$ that the outcome was heads. 
If, however, we are told that the coin is unevenly weighted in favour of tails, then knowing this changes the probability that we assign to the outcomes.  The Bayesian
view leads to probabilities which are constantly being changed as we learn more about the system of interest.  The means of updating our probability distributions is provided by Bayes' rule.  This states that if
we initially assign a probability $p(x)$ that an event $x$ will occur, and then find out $y$, then the new probability we assign to $x$ occurring given
$y$, $p(x|y)$, is given by
\begin{eqnarray}
\label{bayesrule}
p(x|y)&=&\frac{p(y|x)p(x)}{p(y)}=\frac{p(y|x)p(x)}{\int{p(x, y)dx}}\nonumber\\
&=&\frac{p(y|x)p(x)}{\int{p(y|x)p(x)dx}}.
\end{eqnarray}
The Bayesian approach is frequently employed for decision and estimation problems due to the simple way it allows us to include new information.  This approach will now be
used to help us in the problem of estimating the expectation values of two observables for a qubit.

We shall consider performing separate measurements of the observables $\hat A$ and $\hat B$, that is, $\hat A$ will be measured on one copy and $\hat B$ will be measured on a
different copy.  The results of these measurements will be used to update the probability distributions $P(\langle\hat A\rangle )$ and $P(\langle\hat B\rangle )$, which will in turn be used
to estimate the expectation values.  Initially only the results of a measurement of $\hat A$ will be used to update $P(\langle\hat A\rangle )$ and likewise for $P(\langle\hat
B\rangle )$.  We will find that this approach leads to the same total error, equation (\ref{biased}), as we found for biased measurements.  After this we use all the results to update
the probability distributions $P(\cc)$ and $P(\bb)$.  It will be shown that this approach allows us to estimate the expectation values with an error that is lower than equation
(\ref{biased}). 

In the following calculations we will need to perform an average over all possible pure states, where we
assume that the pure states are distributed evenly over the surface of the Bloch sphere.  The average over pure state can be expressed in spherical polar coordinates as it was
in section \ref{separate}.  We shall take ${\bf a}$ 
to point along the $z$ direction of the Bloch sphere, this means that $\langle\hat A\rangle=\langle{\bf a\cdot\hat\sigma}\rangle=\cos(\theta)$.  It should be noted that
the quantities that we will be averaging will be functions of $\langle\hat A\rangle$, and thus will not depend on the angle $\phi$. This means that the average over all pure
states can be written as
\begin{eqnarray}
\frac{1}{4\pi}\int^{\pi}_0{\int^{2\pi}_0{...\sin(\theta)d\theta}d\phi}&=&\frac{1}{2}\int^1_{-1}{...d\cos(\theta)}\nonumber\\
&=&\frac{1}{2}\int^1_{-1}{...d\cc}.
\end{eqnarray}
As we are initially ignorant of the value of $\langle\hat A\rangle$, we shall take the initial probability
distribution of $\cc$ to
be $P(\langle\hat A\rangle )=1/2$ for all values of $\langle\hat A\rangle$.  We then update this probability after performing $N$ measurements of the spin along ${\bf a}$.  As these
measurements are performed on separate but identically prepared systems, updating the probability distribution after all $N$ measurements is equivalent to updating the
probability distribution after each measurement.  From Bayes rule, equation (\ref{bayesrule}), we find that our updated probability is 
\begin{equation}
\label{bayes1}
P(\cc|\{a_j^i\})=\frac{P(\{a_j^i\}|\cc)P(\cc)}{\int{P(\{a_j^i\}|\cc)P(\cc)d\cc}},
\end{equation}
where $\{a_j^i\}$ denotes the $j^{th}$ set of $N$ results for the measurement along ${\bf a}$.  The updated probability distribution can then be used to estimate $\tilde{a}_j$ according to
\begin{equation}
\label{bayesestimate}
\tilde{a}_j=\int^1_{-1}{\cc P(\cc|\{a_j^i\})}d\cc.
\end{equation}
As before the value of the estimate $\tilde{a}_j$ will depend on the measurement results $\{a_j^i\}$.  We can estimate $\langle \hat B\rangle$ by a completely analogous
process, where we have expressions similar to equations (\ref{bayes1}) and (\ref{bayesestimate}).  We proceed by calculating the error that we would obtain for an
estimate of $\langle\hat A\rangle$ for some fixed value of $\cc$ and then we average this result over all possible values of $\cc$.  Thus the total error will be given by
\begin{eqnarray}
\epsilon_T&=&\frac{1}{2}\int^1_{-1}\sum_{\{a_j^i\}}{P_j(\langle\hat A\rangle-{\tilde a_j})^2}d\cc\nonumber\\
&+&\frac{1}{2}\int^1_{-1}\sum_{\{b_k^i\}}{P_k(\langle\hat B\rangle-{\tilde b_k})^2}d\bb,
\end{eqnarray}
where $P_j$ is the probability of obtaining the $j^{th}$ set of results for the $N$ measurements of $\hat A$ and $P_k$ is the probability of obtaining the $k^{th}$ set of results for the
measurement of $\hat B$.  From symmetry the error in the estimate of $\bb$ will be equal to the error in the estimate of $\cc$.   

The results of each measurement of $\hat A$ are independent which means that $P(\{a^i_j\}|\cc)$ will simply be the product of the $N$ single measurement
probabilities $P(a^i_j|\cc)$, where $a^i_j$ refers to the outcome of the $i^{th}$ measurement of $\hat A$.  The probability $P(a^i_j|\hat\rho)$ depends only on the component of the Bloch vector of
$\hat\rho$ along the direction ${\bf a}$, which means that $P(a^i_j|\hat\rho)=P(a^i_j|\cc)$.  For the sake of brevity we shall introduce the notation $I_{n,
\{a_j^i\}}=\int^1_{-1}{\cc^nP(\{a_j^i\}|\cc )d\cc }$, so that $\tilde{a}_j=I_{1,\{a_j^i\}}/I_{0,\{a_j^i\}}$.  This notation allows us to express the error in $\langle\hat A\rangle$ in the compact form
\begin{eqnarray}
\label{comp_error}
\epsilon_a&=&\frac{1}{3}-2\sum_{\{a_j^i\}}{\tilde{a}_j\frac{1}{2}I_{1,\{a_j^i\}}}+\sum_{\{a_j^i\}}{(\tilde{a}_j)^2\frac{1}{2}I_{0,\{a_j^i\}}}\nonumber\\
&=&\frac{1}{3}-\frac{1}{2}\sum_{\{a_j^i\}}{\frac{(I_{1,\{a_j^i\}})^2}{I_{0,\{a_j^i\}}}}.
\end{eqnarray}
Let $r$ denote the number of times spin up was obtained for a set of outcomes $\{a_j^i\}$.  Then we can replace the sum over all possible sets of outcomes in equation (\ref{comp_error}) with a sum over $r$.  This will
lead to a factor $N!/[r!(N-r)!]$ appearing within the summation to take account of the fact that some of the sets $\{a_j^i\}$ are permutations of each other and thus different sets will have the same
value of $r$.  To calculate the error the integrals $I_{0,r}$ and $I_{1,r}$ will need to be evaluated.

By changing the variable of integration to $u=1/2(1+\cc )$, we obtain $I_{0,r}=2\int^1_0{u^r(1-u)^{N-r}}du$, which is of the form of a
beta function $2B(r+1, N-r+1)=2r!(N-r)!/(N+1)!$.  The same change of variable allows us to express $I_{1,r}$ as a linear combination of beta functions, $I_{1,r}=4B(r+2, N-r)-2B(r+1,
N-r+1)$.  It can be shown that 
\begin{eqnarray}
\epsilon_a&=&\frac{1}{3}-\frac{1}{2}\sum_r^N{\frac{N!}{r!(N-r)!}\frac{(I_{1,r})^2}{I_{0,r}}}\nonumber\\
&=&\frac{1}{3}-\sum_r^N{\frac{(4r^2-4Nr+N^2)}{(N+1)(N+2)^2}}\nonumber\\
&=&\frac{2}{3(N+2)}.
\end{eqnarray}
The calculation for the error in $\langle\hat B\rangle$, $\epsilon_b$, is completely analogous.  Hence the total error will be the same as was found in
equation (\ref{biased}) for the biased estimates.  The Bayesian analysis therefore naturally takes account of biasedness.  

As was noted in sections \ref{separate} and \ref{biasing}, a joint measurement can sometimes allow
estimation of two observables with an error lower than equation (\ref{biased}).  It is, however, possible to improve
upon equation (\ref{biased}) while still performing separate measurements of the two observables.  To achieve this we must change the way we update the probabilities $P(\cc)$ and $P(\bb)$.  Instead of using only
the results of measurements of $\hat A$ to update $P(\cc)$ we also include the results of the measurements of $\hat B$.  This means that equation (\ref{bayes1}) changes to 
\begin{equation}
P(\cc|\{a_j^m,b_k^n\})=\frac{P(\{a_j^m,b_k^n\}|\cc)P(\cc)}{\int{P(\{a_j^m,b_k^n\}|\cc)P(\cc)d\cc}},
\end{equation}
where $\{a_j^m,b_k^n\}$ denotes the sets of outcomes $\{a_j^m\}$ and $\{b_k^n\}$, for the measurements of $\hat A$ and $\hat B$.  Similarly the estimate of $\hat A$ changes to $\tilde{a}_j=\int{\cc
P(\cc|\{a_j^m,b_k^n\})d\cc}$.  The measurements of
$\hat A$ and $\hat B$ are performed upon separate copies of the system, thus 
\begin{eqnarray}
P(\{a_j^m,b_k^n\}|\cc)=P(\{a_j^m\}|\cc)P(\{b_k^n\}|\cc)\nonumber\\
=\left[\prod_m^N{P(a_j^m|\cc)}\right]\left[\prod_n^N{P(b_k^n|\cc)}\right].  
\end{eqnarray}
As before
$P(a_j^m|\cc)=P(a_j^m|\hat\rho)$.  It will, however, generally not be true that $P(b_k^n|\cc)=P(b_k^n|\hat\rho)$.  Suppose that $\cc$ assumes the value $x$.  We can then relate $P(b_k^n|\cc=x)$ to
$P(b_k^n|\hat\rho)$ by averaging
$P(b_k^n|\hat\rho)$ over all the pure states $\hat\rho$ for which Tr$(\hat\rho\hat A)=x$.  Such states will be confined to a circle on the surface of the Bloch sphere.  A simple calculation shows that
$P(b_k^n|\cc)=\int P(b_k^n|\rho)d\hat\rho=1/2\pm 1/2({\bf a\cdot b})\cc$, where the $+$ occurs when $b_k^n=1$ and the $-$ occurs when $b_k^n=-1$.  It should also be noted that
the integration is performed over the set of pure states, $\hat\rho$, where Tr$(\hat\rho\hat A)=x$.  For the sake of brevity we introduce the notation 
\begin{equation}
I_{n,\{a_j^m,b_k^n\}}=\int^1_{-1}{\cc^nP(\{a_j^m,b_k^n\}|\cc)d\cc},
\end{equation}
as before it will at times be useful to change this notation to $I_{n,rs}$ where $r$
and $s$ are the number of times ${\bf a}$ and ${\bf b}$ were found to be spin up for a given set of outcomes $\{a_j^m,b_k^n\}$.  In this notation we may express the error in $\langle\hat A\rangle$ as 
\begin{eqnarray}
\label{besterror}
\epsilon_a&=&\frac{1}{3}-2\int\sum_{\{a_j^m,b_k^n\}}{P(\{a_j^m,b_k^n\}|\cc)\tilde{a}_j\cc}d\cc\nonumber\\
&+&\int\sum_{\{a_j^m,b_k^n\}}{P(\{a_j^m, b_k^n\}|\cc)(\tilde{a}_j)^2}d\cc\nonumber \\
&=&\frac{1}{3}-\frac{1}{2}\sum_{\{a_j^m,b_k^n\}}{\frac{(I_{1,\{a_j^m,b_m^k\}})^2}{I_{0,\{a_j^m,b_k^n\}}}}\\
&=&\frac{1}{3}-\frac{1}{2}\sum_{r, s}{\frac{N!}{r!(N-r)!}\frac{N!}{s!(N-s)!}\frac{(I_{1,rs})^2}{I_{0,rs}}}\nonumber.
\end{eqnarray}

\begin{figure}
\center{\includegraphics[width=8cm,height=!]
{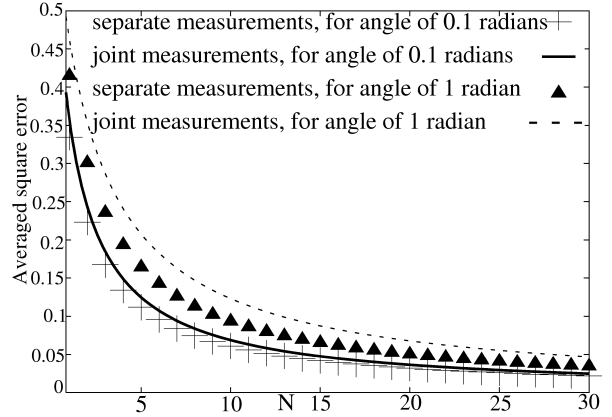}}
\caption{A plot comparing the error for joint and separate measurements, as a function of $N$ for different $\eta$.}
\label{fig2}
\end{figure}

The derivation of $\epsilon_b$ follows a completely analogous procedure, where now we seek $P(\bb|\{a_j^m,b_k^n\})$.  From symmetry it follows that
$\epsilon_b=\epsilon_a$ giving $\epsilon_T=2\epsilon_a$.  The evaluation of the expression (\ref{besterror}) is straightforward for a given value of $N$.  Obtaining an expression for a general
$N$ is however difficult.  Two simple cases exist for which equation (\ref{besterror}) may be readily evaluated.  The first of these is when ${\bf a\cdot b}=0$ meaning that
$P(b^i_j|\cc)=1/2$ for all $i$, and thus $\epsilon_T=4/[3(N+2)]$.  It should
not be surprising that we obtain the same error as was found before in equation (\ref{biased}).  This is because the observables are complementary and thus we would not expect knowledge of the results of
measurements of $\hat B$ to help with estimates of $\langle\hat A\rangle$.  The second case is when ${\bf a\cdot b}=1$, that is, when ${\bf a=b}$.  It can be shown that in this
instance $\epsilon_T=2/[3(N+1)]$, which
is less than equation (\ref{biased}).  The observables $\hat A$ and $\hat B$ are now the same observable and if we wish to estimate a single observable when we have $2N$ copies
of an unknown pure state, then from \cite{Hayashi} we would expect that $\epsilon=1/[3(N+1)]$, which is half of $\epsilon_T=2\epsilon_a=2\epsilon_b$.  The results of equation
(\ref{besterror}) are thus consistent with \cite{Hayashi}, for $\hat A=\hat B$.  Intuitively we would expect that these
two cases should act as upper and lower limits for the total error obtained by this approach.  This intuition is verified by numerical investigations of $\epsilon_T$, plots of
which are shown in figure \ref{fig2}.  It can also be seen that the error that is obtained is always less than the error in equation (\ref{biasJM}), obtained using joint measurements. 
Thus using Bayes' rule allows us to process the measurement results so as to improve the error such that it is no worse and often better than the total error obtained using
the joint measurements described in sections \ref{separate} and \ref{biasing}. 

The Bayesian approach, that we have discussed, leads to an improvement in the error because the results of both observables are used to make our estimates.  It is thus similar
to the approach that was discussed at the end of section \ref{biasing}, which lead to the error given by equation (\ref{simple}).  We shall now compare these two approaches for
${\bf a}\cdot{\bf b}=0$ and ${\bf a}\cdot{\bf b}=1$.  Examining these two situations for equation (\ref{simple}) we find that $\epsilon_T=4/[3(N+2)]$ for ${\bf a}\cdot{\bf b}=0$
and $\epsilon_T=2/[3(N+1)]$ for ${\bf a}\cdot{\bf b}=1$, which is the same as was found for the second Bayesian approach, which led to equation (\ref{besterror}).  For other
values of ${\bf a}\cdot{\bf b}$, and consequently other values of $\eta$, it is found that equation (\ref{simple}) is always greater than the total error obtained numerically
for the Bayesian analysis. 
 
\section{Estimating the expectation values of three complementary spin observables}
\label{3obs}
Thus far we have considered estimating two observables.  We shall now investigate measuring three different observables of a qubit.  The observables that we shall
consider are complementary, for example the $x$, $y$ and $z$ components of spin for a spin 1/2 particle.  As before we shall adopt two different measuring strategies, either measuring each of the
observables separately or performing a joint measurement of
the three observables upon each system.  We assume that we are given $3N$ copies of an unknown state and we again take as our figure of merit the total error, which is averaged
over all the measurement results and all the possible pure states.  Thus 
\begin{eqnarray}
\label{3error}
\epsilon_T=\int\sum_j
P_j[(\langle\hat\sigma_x\rangle-\tilde{x}_j)^2+(\langle\hat\sigma_y\rangle-\tilde{y}_j)^2\nonumber\\
+(\langle\hat\sigma_z\rangle-\tilde{z}_j)^2]d\hat\rho,
\end{eqnarray}
where $\tilde{x}_j$, $\tilde{y}_j$ and $\tilde{z}_j$ are the estimates of $\langle\hat\sigma_x\rangle$, $\langle\hat\sigma_y\rangle$ and $\langle\hat\sigma_z\rangle$
respectively.

We begin by looking at measuring each observable separately.  If we use the unbiased estimates of section \ref{separate} then we find that $\epsilon_T=2/N$.  As before we may
improve upon this by using biased estimates, which allow us to estimate the observables with an error of $\epsilon_T=2/(N+2)$.  Using Bayesian methods will not allow us to
improve upon this error as the three observables are all complementary to each other.  

Our second approach is to perform a joint measurement that gives us information about each
of the three observables.  Again we apply the condition of joint unbiasedness so that we have $\langle\hat\sigma_{x J}\rangle=\alpha\langle\hat\sigma_x\rangle$,
$\langle\hat\sigma_{y J}\rangle=\beta\langle\hat\sigma_y\rangle$ and $\langle\hat\sigma_{z J}\rangle=\gamma\langle\hat\sigma_z\rangle$, where $\langle\hat\sigma_{x J}\rangle$,
$\langle\hat\sigma_{y J}\rangle$ and $\langle\hat\sigma_{z J}\rangle$ are the jointly measured expectation values.  In \cite{busch} a condition similar to the
inequality (\ref{ineq}) is discussed for a joint measurement with sharpnesses $\alpha$, $\beta$ and $\gamma$, which states that $\alpha^2+\beta^2+\gamma^2\le 1$.  One important
difference between this inequality and the inequality (\ref{ineq}) is that in general this inequality is only a sufficient condition for the existence of a joint measurement of three different
spin components.  It is thus possible for a joint measurement of three spin components to exist which does not satisfy this inequality.  A measurement of $\hat\sigma_x$,
$\hat\sigma_y$ and $\hat\sigma_z$ will, however, not violate the inequality \cite{steve}.  A short proof of this fact is given in the appendix.  With this
in mind we shall consider joint measurements that satisfy $\alpha^2+\beta^2+\gamma^2= 1$ to be an optimal joint measurement.  We shall rescale our measurement results from $\pm 1$ to $\pm K/\alpha$, $\pm K/\beta$ and $\pm K/\gamma$ for $x$, $y$ and $z$ respectively, where $K$ is a
constant.  As we are
completely ignorant of the state, symmetry would suggest that we should measure each observable equally well and thus $\alpha=\beta=\gamma=1/\sqrt{3}$.  We find that the total
error is given by
\begin{equation}
\epsilon_T=\frac{3-\alpha^2}{3-\alpha^2+2\alpha^2N}=\frac{4}{4+N}.
\end{equation}
It can be seen that if we perform joint measurements then the total error is greater than the error obtained for the biased separate measurements of the observables.

One interesting point to note is that the joint measurement of $\hat\sigma_x$, $\hat\sigma_y$ and $\hat\sigma_z$ is informationally complete \cite{infocomplete}.  In other words the joint
probability distribution that we obtain will be different for each state.  It is also worth noting that the joint measurement is not a symmetric informationally complete
measurement \cite{sic}.  The reason for this is that some of the POM elements, that describe the joint measurement, are orthogonal to each other while others are not.  In
conventional state tomography we reconstruct the state by performing several different measurements repeatedly, but informationally complete measurements allow us to reconstruct
the state by repeatedly performing only one type of measurement.  For state estimation a useful figure of merit is the trace distance, which is defined as $D(\hat\rho_1,
\hat\rho_2)=Tr|\hat\rho_1-\hat\rho_2|$, where $|\hat O|=\sqrt{\hat O^{\dagger}\hat O}$.  It is easy to show that
for two qubits with Bloch vectors ${\bf c}_1$ and ${\bf c}_2$ the trace distance equals $D(\hat\rho_1, \hat\rho_2)=|{\bf c}_1-{\bf c}_2|$.  If we are given a state $\hat\rho$ and we
estimate it to be $\hat\rho_{est}$ then we could choose our method such that we minimize 
\begin{eqnarray}
D(\hat\rho,\hat\rho_{est})&=&[(\langle\hat\sigma_x\rangle-\tilde{x})^2+(\langle\hat\sigma_y\rangle-\tilde{y})^2\nonumber\\
&+&(\langle\hat\sigma_z\rangle-\tilde{z})^2]^{1/2},
\end{eqnarray}
where $\tilde{x}$, $\tilde{y}$
and $\tilde{z}$ are respectively the $x$, $y$ and $z$ components of the Bloch vector of $\hat\rho_{est}$.  The measurements that we perform to
estimate the state will have different outcomes, which will lead us to obtain a different estimate of the state.  Therefore the distance is averaged over all possible measurement results. 
The state which is being estimated is unknown to us, thus we should perform a second average
over all possible input states.  This figure of merit for state estimation is connected in a simple way to the averaged total error by 
\begin{equation}
\epsilon_T=\int\sum_jP_j D(\hat\rho, \hat\rho_{est}^j)^2d\hat\rho,
\end{equation}
where $\hat\rho_{est}^j$ denotes our estimate of $\hat\rho$ given we obtained the $j^{th}$ set of measurement outcomes.  

In view of our earlier results we find that although
informationally complete joint measurements allow us to estimate a state by performing the same measurement 
repeatedly, they are not as efficient as performing biased separate non-informationally complete measurements of the spin along three orthogonal directions.  While the
informationally complete joint measurement is not as efficient as simply measuring along three orthogonal directions, this does not mean that informationally complete
measurements are not useful for state estimation.  For an example of informationally complete POMs applied to state estimation see \cite{rehacek}, which investigates using four element informationally complete POMs to perform
state tomography on a qubit.  

If our methods for estimating expectation values are used for state estimation, this may lead to an estimated state with a Bloch vector with a
length that is greater than one.  This is a consequence of the figure of merit we have used.  In state estimation, one  may choose to overcome this problem by renormalising the
obtained Bloch vector so that it has unit length.  An alternative approach is to use an estimation procedure, in which it is impossible to produce an estimated state with
a Bloch vector that has length greater than one.  This approach is adopted in \cite{rehacek}.
%

\section{Conclusions}
\label{conc}
We have investigated estimating the expectation values of two different observables for a qubit system that is prepared in an unknown pure state.  Two different approaches were considered, performing separate measurements of each
of the observables or performing a joint measurement of the observables upon each single system.  It was found that measuring the observables separately leads to a smaller total
error in the estimates, if we use either the Bayesian methods or use estimates of the form given by equation (\ref{Estsimple}).  If instead, only the measurement results
relating to a particular observable are used for estimating the expectation value of that observable, then joint measurements may be better.  This is because joint measurements
naturally take account of information which is shared between the observables when these are not complementary. 

The problem with the estimation procedures described in sections \ref{separate} and \ref{biasing} was that estimates of each observable were made using only the information obtained
from measurements of that observable.  When the angle between ${\bf a}$ and ${\bf b}$ is small then the observables share information.  This is exploited by
joint measurements, giving a lower error.  The performance of separate measurements can be improved if we make an estimate of an expectation value based upon the results for both observables.  By using
this approach it was found that separate measurements lead to a smaller error than joint measurements, for all values of the angle between ${\bf a}$ and ${\bf b}$.

We finally examined estimating three complementary spin 1/2 observables of a qubit.  It was found that performing separate measurements of all of the three observables
leads to a smaller error than performing a joint measurement of all three observables.  The situation of measuring three complementary spin components of a qubit is important
for state tomography, as the results of the measurements enables us to determine the state of the qubit.  In view of our findings it can be seen that if we wish to perform state
tomography on a qubit, when we have a limited number of copies, then performing separate measurements of the $x$, $y$ and $z$ spin components is more efficient that performing
an optimal joint measurement of the three observables.

\acknowledgments
We would like to thank Stephen Barnett for his helpful feedback and for allowing us to reproduce his proof that the condition $\alpha^2+\beta^2+\gamma^2\le 1$, is a necessary condition for a
joint measurement of three complementary spin components to be performed.  This proof is reproduced in the appendix.  We would also like to thank John Jeffers for his helpful
feedback.  E.A. acknowledges the Royal Society for financial support. 

\appendix
\section{}
Suppose we wish to perform a joint measurement of $\hat\sigma_x$, $\hat\sigma_y$ and $\hat\sigma_z$.  We shall apply the joint unbiasedness condition so that we have
$\langle\hat\sigma_{x J}\rangle=\alpha\langle\hat\sigma_x\rangle$, $\langle\hat\sigma_{y J}\rangle=\beta\langle\hat\sigma_y\rangle$ and $\langle\hat\sigma_{z
J}\rangle=\gamma\langle\hat\sigma_z\rangle$.  In order to perform this joint measurement there must exist an eight-element POM, $\{\hat\Pi^{xyz}_{i,j,k}\}$, so that we can
assign outcomes to each of the three spin observables.  We define the three marginal POMs, which describe the probability of obtaining a given outcome for each of the three
observables, to be $\hat\Pi^{\alpha x}_{\pm}=\sum_{j,k}{\hat\Pi^{xyz}_{\pm,j, k}}$,  $\hat\Pi^{\beta y}_{\pm}=\sum_{i,k}{\hat\Pi^{xyz}_{i, \pm, k}}$ and $\hat\Pi^{\gamma
z}_{\pm}=\sum_{i,j}{\hat\Pi^{xyz}_{i,j,\pm}}$.  If we assign the results $\pm 1$ to the outcomes of each spin measurement, then for the joint unbiasedness condition to be
satisfied the marginal POMs must have the form \cite{busch} 
\begin{eqnarray}
\label{marginal}
\hat\Pi^{\alpha x}_{\pm}&=&\frac{1}{2}(\hat 1\pm\alpha\hat\sigma_x)\text{,   }\hat\Pi^{\beta y}_{\pm}=\frac{1}{2}(\hat 1\pm\beta\hat\sigma_y)\nonumber\\
\hat\Pi^{\gamma z}_{\pm}&=&\frac{1}{2}(\hat 1\pm\gamma\hat\sigma_z).
\end{eqnarray}

Consider now the situation when we have two qubits prepared in a singlet state 
\begin{equation}
|\psi\rangle_{12}=\frac{1}{\sqrt{2}}(|+-\rangle_{12}-|-+\rangle_{12}),
\end{equation}
where $\hat\sigma_x|\pm\rangle=\pm|\pm\rangle$. 
Suppose the joint measurement $\hat\Pi^{xyz}_{i,j,k}$ is performed on the first qubit and we obtain the outcome $+++$, that is all three spin components are found to be spin up. 
Given this outcome the second qubit will now be prepared in the state $\hat\rho_2$.   

If an unsharp measurement of $\hat\sigma_x$ were performed on the first qubit of $|\psi\rangle$ and a projective measurement of $\hat\sigma_x$ were performed on the second qubit, then the probability
distribution for the outcomes would be $P_{12}(\pm,\pm)=1/4(1-\alpha)$ and $P_{12}(\pm,\mp)=1/4(1+\alpha)$.  The conditional probability for the
results of the measurement on the second qubit, given that we obtained `spin up' for the unsharp measurement on the first qubit, are found to be  
\begin{equation}
P_{21}(\pm|+)=\frac{1}{2}\mp\frac{\alpha}{2}.
\end{equation}
A joint measurement of the $x$, $y$ and $z$ spin components will implement three unsharp spin measurements described by the marginal POMs (\ref{marginal}).  If we obtain the
outcome $+++$ and thus prepare the second qubit in the state $\hat\rho_2=1/2(\hat 1+{\bf c}\cdot\hat\sigma)$, then the probability of obtaining the the result $\pm$ for a
projective measurement of $\hat\sigma_x$ will be $\langle\pm|\hat\rho_2|\pm\rangle=(1\pm c_x)/2$.  If the joint measurement does indeed implement an unsharp
measurement of $\hat\sigma_x$, with sharpness $\alpha$, then $P_{21}(\pm|+)=(1\mp\alpha)/2=\langle\pm|\hat\rho_2|\pm\rangle$.  From this we obtain that $c_x=-\alpha$.  Using
similar arguments we can find that $c_y=-\beta$, and $c_z=-\gamma$.  A necessary condition for $\hat\rho_2$ to be a valid density operator is that $|{\bf c}|^2\le 1$, which
implies that  
\begin{equation}
\label{necessary}
\alpha^2+\beta^2+\gamma^2\le 1.
\end{equation}
We have thus established that (\ref{necessary}) is a necessary condition for us to be able to perform a joint measurement of $\hat\sigma_x$, $\hat\sigma_y$ and $\hat\sigma_z$,
with sharpnesses of $\alpha$, $\beta$ and $\gamma$, respectively.

\end{document}